\documentclass[aps,prl,10pt,twocolumn,groupedaddress,notitlepage,showpacs,floatfix]{revtex4-1}
\pdfoutput=1
\usepackage{graphicx,graphics,epsfig,subfigure,times,bm,bbm,amssymb,amsmath,amsfonts,amsthm,mathrsfs,MnSymbol}
\usepackage{gensymb}
\usepackage[matrix,frame,arrow]{xypic}
\usepackage[pdfstartview=FitH]{hyperref}
\usepackage{subfigure}
\hypersetup{
    colorlinks=true,       
    linkcolor=red,          
   citecolor=magenta,        
    filecolor=magenta,      
    urlcolor=cyan,           
    runcolor=cyan
}
\usepackage[pdftex]{color}
\usepackage{braket}
\usepackage{enumerate}
\usepackage[normalem]{ulem}
\usepackage[usenames,dvipsnames]{xcolor}
\usepackage{multirow}
\usepackage{mathtools}

\definecolor{orange}{rgb}{1,0.5,0}

\newcommand{\RN}[1]{\textup{\lowercase\expandafter{\romannumeral#1}}}

\newcommand{\bes} {\begin{subequations}}
\newcommand{\ees} {\end{subequations}}
\newcommand{\bea} {\begin{eqnarray}}
\newcommand{\eea} {\end{eqnarray}}

\definecolor{gold}{rgb}{0.85,.66,0}

\newcommand{\norm}[1]{\ensuremath{\left\|#1\right\|}} 

\newcommand{\beq}{\begin{equation}}

\newcommand{\eeq}{\end{equation}}

\newcommand{\ignore}[1]{}
\newcommand{\mc}[1]{\mathcal{#1}}

\def\bS{\textbf{S}}

\def\tr{\mathrm{Tr}}

\def\s{\sigma}

\def\>{\rangle}
\def\<{\langle}

\def\s0{I}

\newcommand{\ig}[1]{}

\def\dgr{\dagger}

\begin{document}

\title{Propagation of correlations in Local Random Quantum Circuits.}
\author{Siddhartha Santra and Radhakrishnan Balu}
\affiliation{ U.S. Army Research Laboratory, Computational and Information Sciences Directorate, Adelphi, Maryland, U.S.A. - 20783\\\& U.S. Army Research Laboratory, Computational and Information Sciences
Directorate, ATTN: CIH-N, Aberdeen Proving Ground, Maryland, U.S.A. - 21005-5069}

\begin{abstract}
We derive a dynamical bound on the propagation of correlations in local random quantum circuits - lattice spin systems where piecewise quantum operations - in space and time - occur with classical probabilities. Correlations are quantified by the Frobenius norm of the commutator of two positive operators acting on space-like separated local Hilbert spaces. For times $t=O(L)$ correlations spread to distances $\mc{D}=t$ growing, at best, diffusively for any distance within that radius with extensively suppressed distance dependent corrections whereas for $t=o(L^2)$ all parts of the system get almost equally correlated with exponentially suppressed distance dependent corrections and approach the maximum amount of correlations that may be established asymptotically.
\end{abstract}
\maketitle


\begin{figure}[t]
 \centering
\includegraphics[width=.8\columnwidth,height=5cm]{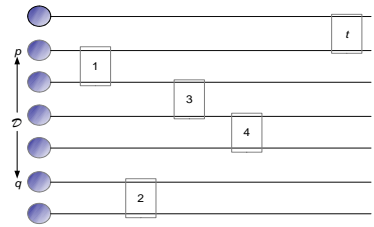}
 \caption{(color online) Schematic action of a LRQC defined on circular chain of qudits (blue circles, top qudit interacts with the bottom one) with nearest neighbor connectivity. At discrete time intervals a single 2-body geometrically local random unitary acts on the chain. The circular topology makes the action of the LRQC translationally invariant over all the vertices of the graph.}
  \label{uniformpathgraph}
\end{figure}

\noindent{\emph{Introduction}}.- In ensembles of quantum systems evolving in time under random local unitary interactions how do correlations of observable measurements spread \emph{on an average} ? For \emph{individual} quantum systems on discrete lattices under local Hamiltonian dynamics  the answer is provided in the form of the Lieb Robinson bound \cite{liebrobin:1972}, and it's generalizations \cite{burrell:2007,poulin:2010}, that establish a dynamical bound on the operator norm of space-like separated observables under dynamics governed by a local Hamiltonian. With the concept of \emph{typicality} \cite{goldstein:2010} - properties that hold statistically on an average - playing an increasingly important role in explaining foundational as well as operational questions in quantum statistical mechanics such as quantum equilibration \cite{goldstein:2006}, thermodynamics \cite{gemmerbook}, area laws for entanglement entropy \cite{hamma:2012a} etc. it is important to have an estimate, in relevant families of quantum systems, for an average dynamical bound on the evolution of correlations in typical states produced via such random processes.

Here we obtain a dynamical bound on the average of Frobenius norm of the commutator of operators in a natural model of a Local Random Quantum Circuit (LRQC). These are stochastic circuits comprised of unitaries with finite support that respect the local-interaction structure of the defining underlying graph representing the constituents of a quantum many body system \cite{hamma:2012a,zanardi:2013a,oliveira:2007,dahlsten:2007}. LRQCs are useful as tools to formulate statistical statements about ensembles of physical systems governed by random local interactions and have been studied to model efficiency and typicality of entanglement generation in random two-party processes \cite{oliveira:2007}, entanglement dynamics of time dependent local hamiltonians \cite{hamma:2012a}, equilibration in quantum systems, as approximate polynomial designs \cite{harrow:2009,brandao:2012}, to define quantum error-correcting codes and quite generally in the decoupling approach to quantum information \cite{brown:2013}. 

\begin{figure} 
\centering
\includegraphics[width=\columnwidth,height=5cm]{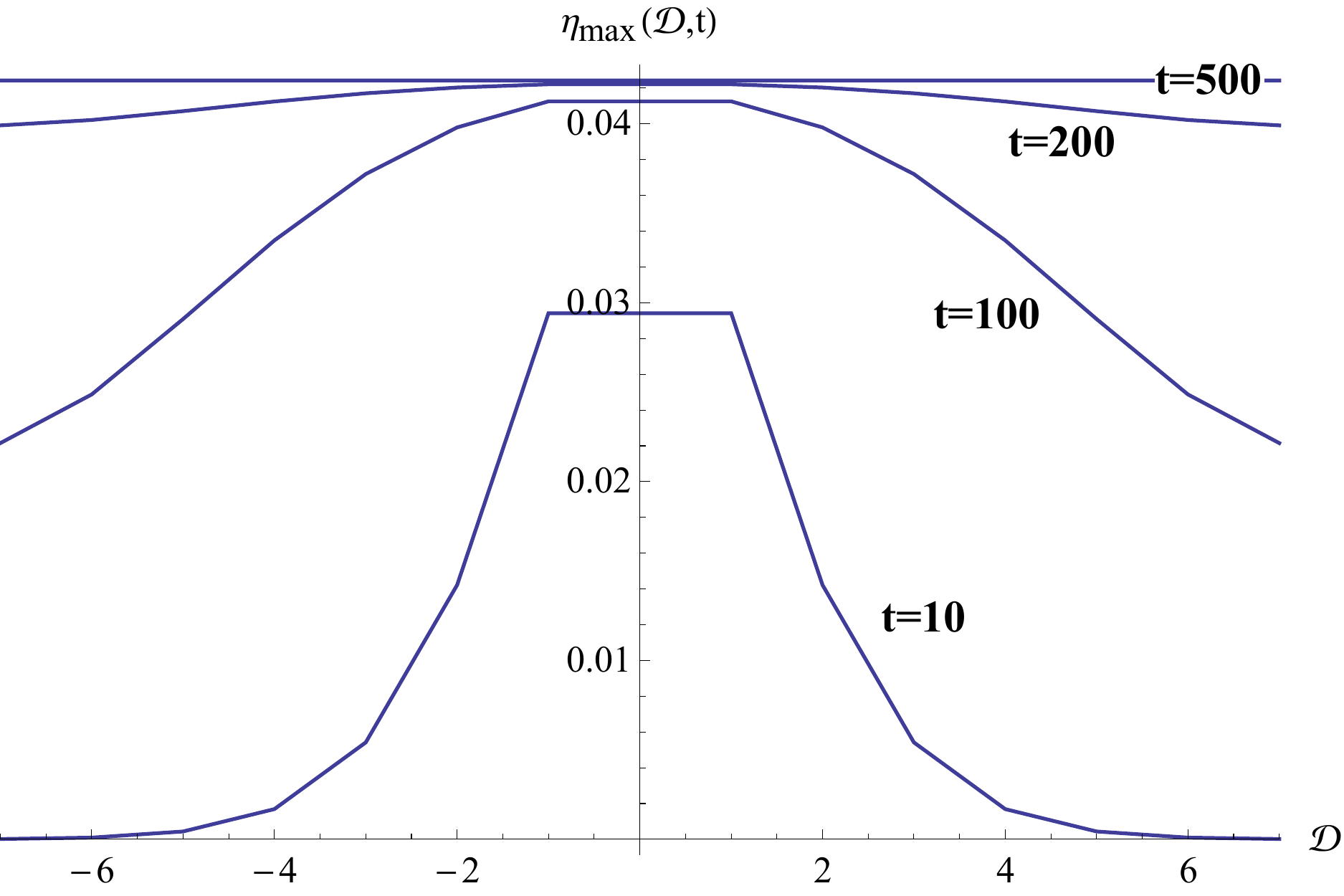}
\caption{(color online) Dynamical evolution of the upper bound $\eta_{\text{max}}(\mc{D},t)\geq\eta(\mc{D},t)$ on correlations in a circular chain of 15 spins (scaled by $d^{L/2}$) for positive operators $\hat{O}_p=\{\{.5,0\},\{0,.3\}\},\hat{O}_q=\{\{.7,0\},\{0,.1\}\}$ at 4 different $t=10,100,200,500$ time steps of the LRQC. The first spin is chosen at the site $p=0$ on the chain which has 7 sites each on either side. The negative and the positive parts of the x-axis represent these two sides of the chain. The data is obtained by numerical evaluation of the R.H.S in ineq.~(\ref{r1r2bound}). Asymptotically the bound approaches the value $\mc{M}=.0424$.}
\label{circchainL15}
\end{figure}

In our model of a LRQC on a spin lattice, Fig.~(\ref{uniformpathgraph}), at each discrete-time step, two random processes occur in succession. First, a support for interactions is chosen according to some distribution over the possible local interactions, followed by a choice of a unitary operator according to another probability distribution over unitary operators with the same support. We show that \emph{on an average} over the circuit realizations, correlations spread at the rate of the size of the support of local unitaries per time step resulting in a strictly linear light cone. Within this light cone we determine a bound on the dynamical correlation in terms of the parameters of the two probability distributions defining the model - namely the probability distribution over the edges and the distribution over local unitaries, Fig.~(\ref{circchainL15}). We establish two distinct time regimes, one for times of the order of system size when average (over circuits) root mean (over basis states) square correlations grow diffusively for distances within the light cone and the other for times greater than the square of the system size where all parts of the system are almost equally correlated and close to the maximum achievable, which value we establish.\\

\noindent{\emph{Local Random Quantum Circuits}}.-
A local random quantum circuit $C^t[\mc{L},\Xi,q^{(t)}]$, of depth $t$ on a graph $\mc{G}=(V,E)$, with an associated Hilbert space $\mathcal{H}_V=\otimes_{i\in V}\mathcal{H}_i$, is specified by three quantities: local regions where interactions may take place, a distribution over unitaries acting on those local regions and a rule assigning probability weights to sequences of local regions chosen in $t$ time steps. Mathematically one specifies subsets `$\mc{L}$' of the power set $2^V$ with elements $S\in \mc{L}$ where unitary operations may take place, probability density $\Xi=\{d\mu_S\}$ over the group of unitaries $\mc{U}_S$ acting on those local regions $S\in \mc{L}$, and a probability law $q^{(t)}:\mc{L}^t\mapsto [0,1]$, assigning sequences of local regions of length $t$ a probability weight. Given this data, a LRQC of depth $t$ is essentially a unitary $U_{\textbf{S}}$ which is a time ordered product of  unitaries acting on the local regions i.e. $U_{\textbf{S}}:=U_tU_{t-1}...U_1$,
where $\textbf{S}=(S_t,S_{t-1},..,2,1)\in\mc{L}^t$ is the sequence of local regions and 
 $U_i\in\mc{U}(S_i) \forall i\in[1,t]$. Thus $U_{\bS}$ is a unitary-valued random variable distributed according to the law $q^{(t)}(\bS)dU_{\bS}=q^{(t)}(\bS)\prod _{i=1}^t d\mu_{S_i}$. The statistics ($n$-th order moment) of any observable $\hat{O}\in \mc{B}(V)$ over the ensemble of circuits with fixed data $\mc{L},\Xi,q^{(t)}$,  where $\mc{B}(V)$ is the algebra of bounded linear operators on $\mc{H}_V$, may be obtained using a set of completely positive trace preserving maps $\mc{R}_n:\mc{B}(V)^{\otimes n}\mapsto\mc{B}(V)^{\otimes n}$ \cite{zanardi:2013a} although for the purpose here we need only the first two such ensemble maps $\mc{R}_1:\mc{B}(V)\mapsto\mc{B}(V)$ and $\mc{R}_2:\mc{B}(V)^{\otimes 2}\mapsto\mc{B}(V)^{\otimes 2}$. These are the averaging superoperator and the second-moment superoperator respectively and have the following actions,
\begin{align}
 \mc{R}_1^t(\hat{O}_1)&=\sum_{\bS\in\mc{L}^t}q^{(t)}(\bS)\int dU_{\bS}U^\dgr_\bS ~\hat{O}_1~U_\bS,\\
\mc{R}_2^t(\hat{O}_1^{\otimes2})&=\sum_{\bS\in\mc{L}^t}q^{(t)}(\bS)\int dU_{\bS}(U^\dgr_\bS)^{\otimes 2} ~\hat{O}_1^{\otimes2}~(U_\bS)^{\otimes 2},
\label{ensemblemaps}
\end{align}
i.e. they average, over sequences of local regions, the operator first and second moments w.r.t. a distribution of unitaries over those local regions. In our model of a LRQC, first a pair of nearest neighbor qudits $a,b\in V$ is chosen uniformly on the graph $\mc{G}(V,E)$ of a circular chain with $V=\{1,2,...,L\},~E=\{(1,2),(2,3),(3,4),..,(L-1,L),(L,1)\}$. Thus the total Hilbert space of our model is $\mc{H}_V=\otimes_{i\in V}\mc{H}_i,\text{Dim}(\mc{H}_i)=d\forall~i\in[1,L]$. The local regions are the edges of $\mc{G}$ which are picked uniformly with a probability $1/L$ in each time step. This is followed by a unitary acting on those qudits chosen randomly according to the Haar measure,  Fig.~(\ref{uniformpathgraph}). The translational invariance of the model due to the circular topology makes the relevant superoperators symmetric in a way that makes the analytical presentation easier.\\

\emph{Dynamical correlations in our model.}-  In the Heisenberg picture the dynamics generated by the concatenation of these two random processes after time $t$ maps any operator $\hat{O}_p$ with support $p\subset V$ to $C^t(\hat{O}_p)=U^\dgr_{\bS}\hat{O}_pU_{\bS}$ that has support $p\cup\bS$. This spread of support due to the LRQC dynamics leads to the building up of correlations between measurements of observables $\hat{O}_p,\hat{O}_q$ at $p$ and any other region $q\subset V$ with time $t$ that depends on the lattice distance $\mc{D}$ between $p$ and $q$, Fig.~(\ref{circchainL15}). For simplicity we choose positive operators \footnote{The linear lightcone nature is not dependent on the sign of the operator spectra but establishing the upper bound is easier with operators with positive spectrum.} with point supports $p,q\in V$ and derive an upper bound on the average over the circuits (denoted by the overbar), $\eta(\mc{D},t)=\overline{||[C^t(\hat{O}_p),\hat{O}_q]||_2}$, of the commutator Frobenius norm i.e.,
\begin{align}
\eta(\mc{D},t)&=\overline{\sqrt{\tr\{[C^t(\hat{O}_p),\hat{O}_q]^\dgr[C^t(\hat{O}_p),\hat{O}_q]\}}}.
\label{eq:eta}
\end{align}
Note that the square of the Frobenius norm of a commutator for two observables $A,B$ with spectral resolutions $A=\sum_i a_i\ket{a_i}\bra{a_i},B=\sum_i b_i\ket{b_i}\bra{b_i}$ is given by,
\begin{align}
\tr\{[A,B]^\dgr[A,B]\}=\sum_{i,j}(a_i-a_j)^2|\braket{a_i|B|a_j}|^2,
\label{frobdef}
\end{align}
which sums the modulus square of matrix elements of $B$ linking two eigenstates of $A$ weighted by the square of the difference of the corresponding eigenvalues - quantifying the sum of the square of correlations in measurements of $A$ and $B$ over a complete basis. Normalizing first the quantity in (\ref{frobdef}) by the dimension of the space and then taking the square root gives the root mean square (RMS) correlation along any basis state of the space. The quantity of interest, for us, is thus (\ref{eq:eta}) normalized by $d^{L/2}$ (root of the dimension of $\mc{H}_V$) i.e. $\eta(\mc{D},t)/d^{L/2}$ which gives the average (over the circuits) RMS (over the Hilbert space) correlations due to the stochastic unitary dynamics of LRQCs. We note that  the Frobenius norm also provides a (weak) upper bound on the operator norm \footnote{For bounded linear operators the norms obey $\norm{\hat{O}}_\alpha\geq\norm{\hat{O}}_\beta~\forall~\alpha\leq\beta$ and thus the commutator two-norm upper bounds its infinite norm $\norm{[\hat{O}_p,\hat{O}_q]}_\infty\leq\norm{[\hat{O}_p,\hat{O}_q]}_2$.}.\\
\begin{figure} 
\centering
\includegraphics[width=\columnwidth]{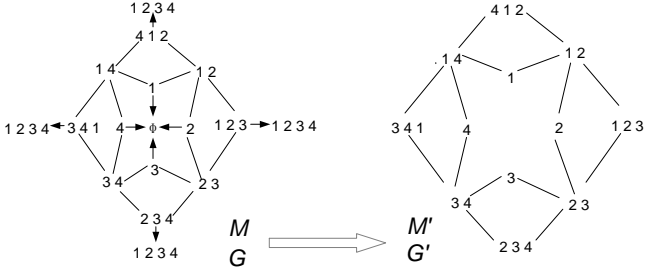}
\caption{(color online) Graph derived from adjacency structure of $M$ - matrix representation for $\mc{R}_2$ on a circular chain with $L=4$ qudits. The vertices of this derived graph show the subsets $S\in W$ over which the swaps $\hat{\mc{T}}_S$ in the invariant subspace for $\mc{R}_2$ have support. The self loops at all vertices (due to diagonal terms in $M$) are not shown in the diagram. By following the evolution of coefficients $c^t_{{1,2}_S}$ under iterations of $M$ we estimate $a_{1,2}(S,t)$.} 
\label{ampflow}
\end{figure}

\emph{Outline of the derivation.}- First, $\eta(t)$ is bound by a functional of iterated ensemble maps $\mathcal{R}_1^t$ and $\mathcal{R}_2^t$ acting on $\hat{O}_p^2$ and $\hat{O}_p^{\otimes2}$ respectively. Next, we show that $\mathcal{R}_1^t(\hat{O}_p^2)$ is independent of $t$ and the entire spatio-temporal dependence of the commutator two-norm is contained in $\mathcal{R}_2^t(\hat{O}_p^{\otimes 2})$. This dependence is analysed using a matrix representation $M$ for $\mc{R}_2$ in an operator basis consisting of swap operators $\hat{\mathcal{T}}_S$  \footnote{Operators $\hat{\mc{T}}_S:\mc{B}(S)^{\otimes2}\mapsto\mc{B}(S)^{\otimes2}, S\subset V$ act on basis states of a doubled Hilbert space $\mc{H}_S\otimes \mc{H}_{S'},\mc{H}_S\cong\mc{H}_{S'}$ as $\hat{\mc{T}}_S\ket{\psi_1}\ket{\psi_2}=\ket{\psi_2}\ket{\psi_1}$ and square to the identity $\hat{\mc{T}}_S^2=\openone_S$.} on all subsets $S\subseteq V$ - leading to an algebra over subsets of $V$. Finally, we estimate the matrix entries $M^t_{i,j}$ by considering the symmetry of a graph derived from the adjacency matrix of $M$ with the weights on edges representing the one-step matrix action. This yields all ingredients for evaluating an upper bound on $\eta(t)$.\\

\emph{Bounding the commutator norm.}- We start by using the concavity of the square root function to take the average in Eq.~(\ref{eq:eta}) under the square-root implying,
\begin{align}
&\eta(t)\leq\sqrt{\overline{\tr{\{[C^t(\hat{O}_p),\hat{O}_q]^\dagger [C^t(\hat{O}_p),\hat{O}_q]}\}}}\nonumber\\
&=\sqrt{2\tr(\mathcal{R}_1^t(\hat{O}_p^2)\hat{O}_q^2)-2\tr(\mathcal{R}_2^t(\hat{O}_p^{\otimes2})\hat{O}_q^{\otimes 2}\hat{\mathcal{T}}_V)}
\label{r1r2bound}
\end{align}
which expresses the dynamical bound on the average commutator two-norm as the square root of a linear combination of traces w.r.t. $\hat{O}_q^2$ and $\hat{O}_q^{\otimes2}\hat{\mc{T}}_V$ of the images of $\hat{O}_p^2$ and $\hat{O}_p^{\otimes2}$ under the ensemble maps $\mc{R}_1$ and $\mc{R}_2$ respectively (see supp. mat.). It turns out that
 $\tr(\mathcal{R}_1^t(\hat{O}_p^2)\hat{O}_q^2)=\tr\{\hat{O}_p^2\}\tr\{\hat{O}_p^2\}d^{L-2}$
 is independent of time (see supp. mat.) while $\tr(\mathcal{R}_2^t(\hat{O}_p^{\otimes2})\hat{O}_q^{\otimes 2}\hat{\mathcal{T}}_V)=:\braket{\mathcal{R}_2^t(\hat{O}_p^{\otimes2}),\hat{O}_q^{\otimes 2}\hat{\mathcal{T}}_V}$ decreases monotonically with $t$ eventually reaching an asymptotic value for large times.
To show this we need $\mc{R}_2^t(\hat{O}_p^{\otimes 2})$ which can be obtained by following the iterations of $\mc{R}_2(\hat{O}_p^{\otimes 2})$ for $t$ time steps (see supp. mat.) and requires evaluation of $\mc{R}_2^{n}(\hat{\mc{T}}_{p,p+1})$ and $\mc{R}_2^{n}(\hat{\mc{T}}_{p,p-1})$ for $n=1,2,..,(t-1)$ i.e. iterated $\mc{R}_2$ maps on swap operators $\hat{\mc{T}}_{p,p\pm1}$ supported on nearest neighbors of node $p$. The iterated version gives,
\begin{align}
\mc{R}_2^{t}(\hat{O}_p^{\otimes 2})&=r^{t}\hat{O}_p^{\otimes 2}+\frac{2A}{L}\frac{1-r^t}{1-r}\openone_V\nonumber\\
&~~~~~~~~~~~~+\frac{B}{L}\sum_{S\in W}\{a_1(S,t)+a_2(S,t)\}\hat{\mc{T}}_{S}
\label{hderiv}
\end{align}
where $r=(L-2)/L$ is the probability with which $\mc{R}_2$ leaves $\hat{O}_p^{\otimes 2}$ invariant. $A=\tr(\hat{O}_p^2)(x^2d^3-1)/d(d^4-1)$ and $B=\tr(\hat{O}^2_p)(d-x^2)/(d^4-1)$ are constants (w.r.t. size $L$) that depend only the norm properties of $\hat{O}_p$ through $x=||\hat{O}_p||_1/||\hat{O}_p||_2$.     In Eq.~(\ref{hderiv}) we have denoted by $W\subset 2^V$, the set with elements that are sets of contiguous vertex labels in $V$ i.e. $W=\{S\subset V|S=\{\phi\},\{1\},\{1,2\},...,\{2\},\{2,3\},\{2,3,4\},\{1,2,3,...,L\},\text{etc.}\}$.
As the set of swaps $\hat{\mc{T}}_{S}$ supported on $W$ forms an invariant subspace under $\mc{R}_2$ we expand $\mc{R}_2^t(\hat{\mc{T}}_{p,p-1})=\sum_{S\in W}c^{(t)}_{1_S}\hat{\mc{T}}_S=\sum_{S\in W}(M^t\ket{c^0_1})_S\hat{\mc{T}}_S$, where $M$ is the matrix representation of $\mc{R}_2$ in the basis $T_W:=\{\hat{\mc{T}}_S|S\in W\}$ with $|T_W|=L(L-1)+2$ and $\ket{c^0_1}$ is a column vector with unit entry at the position corresponding to $\hat{\mc{T}}_{p,p-1}$ and zeros elsewhere in the ordered basis for $M$ and similarly for $\mc{R}_2^t(\hat{\mc{T}}_{p,p+1})=\sum_{S\in W}c^{(t)}_{2_S}\hat{\mc{T}}_S$. The expansion coefficients $a_{1,2}(S,t)=\sum_{l=0}^{t-1}c^{(t-1-l)}_{{1,2}_S}r^l\geq0$ are obtained by evaluating $c^{(n)}_{{1,2}_S}=(M^{n}\ket{c^0_{1,2}})_S,n\in[0,t-1]$, that in general is difficult to obtain exactly \footnote{The spectral gap of the second moment operator $M$ for some RQC models have been studied in the literature however here we need estimates for the individual matrix entries in $M^n$.}. However a lower estimate for $a_{1,2}(S,t)$ is enough to lower bound Eq.~(\ref{hderiv}) and in turn upper bound Eq.~(\ref{r1r2bound}).

We estimate the elements of $M^n$ using a graph $G(W,E')$, derived from the adjacency structure of $M$ whose vertices are the sets of supports of operators that form the basis for $M$ and whose edges are pairs of elements in $W$ with corresponding non-zero entry in $M$, Fig.~(\ref{ampflow}). This derived graph shows the dependence of the coefficients $c^{(t)}_{{1,2}_S}$ at time $t$ on those at the previous time step as $c^{(t)}_{{1,2}_S}\forall S\in W$ depends only on those $c^{(t-1)}_{{1,2}_{S'}}$ for which $S,S'$ are connected by an edge. Starting from $\ket{c^0_{1,2}}_S=\delta(S,S_L=\{p,p-1\})+\delta(S,S_R=\{p,p+1\})$ i.e. unity at fiducial nodes $S_L,S_R$, corresponding to the operators $\hat{\mc{T}}_{p,p-1}$ and $\hat{\mc{T}}_{p,p+1}$, applying $M$ transfers weights to other nodes $S'\in W$. All these weights eventually flow to the sink nodes $S=\{\phi\},\{V\}$, that span the unit eigenspace for $M$ \cite{zanardi:2013a}. Removing the sink nodes (and the edges including these nodes) from $G$ amounts to omitting from $M$ the rows and columns corresponding to $S=\{\phi\},\{V\}$ yielding a positive, symmetric matrix $M'$ with a corresponding graph $G'(W\backslash\{\{\phi\},\{V\}\},E'')$. This is useful since $M'$ has only two kinds of entries:  $r=(L-2)/L$ on the diagonal (self loops in $G'$) while $u=N_d/L=d/(d^2+1)L$ for the other non-zero off-diagonal entries (edges in $G'$). Then it can be shown that $c_{{1}_s}^{(n)}=({M'}^n)_{S,S_L}\geq \Theta(t-D_{i,j})\binom{t}{D_{ij}}u^{D_{ij}}r^{t-D_{ij}}$ and $c_{2_s}^{(n)}=({M'}^n)_{S,S_R}\geq \Theta(t-D_{i,j})\binom{t}{D_{ij}}u^{D_{ij}}r^{t-D_{ij}}$ with $D_{ij}=\text{Dist}(S_i,S_j)$ the graph theoretic distance between nodes labeled by $S_i,S_j$ on the derived graph $G'$ (different from the distance $\mc{D}$ between nodes $p,q$ on the circular chain) resulting in $a_{1,2}(S,t)\geq{u^Dr^{t-1-D}t^{D+1}}/{(D+1)^{D+1}}=a_{1,2}^*(S,t)$ for the non-sink nodes while the same for the sink nodes is given by $a_{1,2}(S,t)=2u\sum_{\text{Dist}(S',S)=1}\sum_{n=1}^{t-1}a_{1,2}(S,n)$ i.e. $2u$ times the sum over nearest neighbors of the sinks of their coefficient values for all previous times. 

Further, using Eq.~(\ref{hderiv}) to obtain $\braket{\mathcal{R}_2^t(\hat{O}_p^{\otimes2}),\hat{O}_q^{\otimes 2}\hat{\mathcal{T}}_V}$ requires evaluation of $Q_s=\braket{\hat{\mc{T}}_S,\hat{O}_q^{\otimes 2}\hat{\mc{T}}_V}$ which value depends on whether the set $S$ includes the node labelled $q$ or not. Indeed we have (with $|S|$ the size of the subset) $Q_S=d^{L+|S|-2}\tr(\hat{O}_q)^2=Q_{1_S}$ for $q\cap\{S\}=q$ and $Q_S=d^{L+|S|-1}\tr(\hat{O}^2_q)=Q_{2_S}$ for $q\cap\{S\}=\phi$, which let's us write (where $a(S,t)=a_{1}(S,t) \text{~or~}a_2(S,t)$), 
\begin{align}
&\sum_{S\in W}a(S,t)Q_S=\sum_{S\in W}a(S,t)Q_{2_S}-\sum_{\overset{S\in W}{S\cap\{q\}=q}}a(S,t)\Delta_S
\label{spatdepform}
\end{align}
with $a(S,t),Q_{2_S},Q_{1_S},\Delta_S=(Q_{2_S}-Q_{1_S})\geq 0$. Together with Eq.~(\ref{r1r2bound}) this implies that to upper bound $\eta(\mc{D},t)$ we need the lower bound on the first term and an upper bound on the second term in the R.H.S. of Eq.~(\ref{spatdepform}) and the latter contains the spatial dependence in the bound for $\eta(\mc{D},t)$ because it sums over only those labels of sets that have non-null intersection with $q$. Analytically we obtain these bounds for short times when $t=O(L)$ and long times $t=o(L^2)$ where the fiducial and sink nodes have the largest coefficients $a(S,t)$ respectively (see supp. mat.).

\emph{Short time regime}.- For $t\leq T_1=(1+1/d^2)(L-2)\sim O(L)$ most of the weight in the expansion for $\mc{R}_2^t(\hat{\mc{T}}_{p,p\pm1})$ is associated with the fiducial nodes as $a(S,t)\sim (ut/D)^D$ falls exponentially with distance $D$ (on the derived graphs) since $ut<1$. We thus keep only the fiducial node contributions for the first term in Eq.~(\ref{spatdepform}) and upper bound the second term using contributions from the nodes $S, S\cap q=q$ nearest to $S_{L,R}$ each giving the dynamical bound,
\begin{align}
&\frac{\eta(\mc{D},t)}{\mathcal{M}d^{L/2}}\leq\Theta(t-\mc{D})[m_1(1-r^t)-m_2r^t\frac{udt}{r}+m_3r^t(\frac{udt}{r\mc{D}})^{\mathcal{D}}]^{1/2}\nonumber\\
\label{smalltform1}
\end{align}
where $\mc{M}=(\{2(d-x^2)(d-y^2)\}^{1/2}/d^2)||\hat{O}_p||_2||\hat{O}_q||_2$ is the asymptotic value $\eta(\mc{D},t\to\infty)$, and with $m=(d-x^2)(d-y^2)/d^2$, $m_1=\{d^3(d-x^2)/(d^4-1)\}/m$ is a number that depends only on the properties of operators $\hat{O}_p,\hat{O}_q$ whereas $m_2=\{(2d(d-x^2))/m(d^2-1)\}k_1(\frac{udt}{r}),m_3=\{2(d-x^2)(d-y^2)/m(d^2-1)\}k_2(\frac{udt}{r})$ where $k_1(\frac{udt}{r})\geq1$ and $k_2(\frac{udt}{r})\leq1$ are quantities that include higher order contributions in $udt/r=\{d^2/(d^2+1)(L-2)\}t\leq1$. The bound (\ref{smalltform1}) implies a linear increase in the spread of the correlations to distances $\mc{D}=t$ due to the step function $\Theta(t-\mc{D})$. For $t<<L$ the rate of correlation growth for any $\mc{D}\leq t$ is essentially determined by $(1-r^t)$ which is the probability of the map $\mc{R}_2$ acting non-trivially on $\hat{O}_p^{\otimes2}$ in $t$-time steps. Since the second and third terms under the root in bound (\ref{smalltform1}) are extensively suppressed due to the factor of $u=d/(d^2+1)L$, for $1\ll \mc{D}<t\ll L$ only the first term contributes significantly, and one has $\eta(\mc{D},t)/d^{L/2}<_{L\to\infty}\Theta(t-\mc{D})\mc{M}\sqrt{(1-r)m_1}t^{1/2}+\mc{O}(1/L^{\mc{D}/2})$ impying a diffusive growth for the bound on average RMS correlations within the linear light cone. The square of this quantity gives the mean square correlation (not RMS) implying a ballistic growth of the same in this spatio-temporal regime - consistent with the 2-R\'{e}nyi entropy growth \cite{zanardi:2013a}. The time-scale $T_1$ thus estimates the diffusive regime for the growth of average RMS correlations.

\emph{Long time regime}.- For $t\geq T_2=\frac{e(L+1)(L-2)(d^2+1)}{d}\sim o(L^2)$ the maximum contribution to $a(S,t)$ for any particular $S$ comes from the longest paths from the fiducial nodes $S_L,S_R$ to $S$ and the most significant $a(S,t)$ in this time regime are for the two sink nodes $S=\phi,\{1,2,...,L\}$ which flatten out close to an asymptotic value leading to the bound,
\begin{align}
&\frac{\eta(\mc{D},t)}{\mathcal{M}d^{L/2}}\leq[m_1(1-r^t)-h_1f(t)+h_2g(t,\mc{D})]^{1/2}
\label{largetform1}
\end{align}
where $\mc{M},m_1$ are as before and $h_1=d^3(d-x^2)/m(d^4-1)(d^2+1)^2$ depends only on the properties of $\hat{O}_p,\hat{O}_q$. $f(t)$ is a monotonically increasing function of time flattening out at a value of 1 i.e $f(t\to\infty)=0$ and the term $h_2g(t,\mc{D})$ is an exponentially suppressed distance and time dependent term which goes to zero asymptotically i.e. $h_2g(t\to\infty,\mc{D})=0$ \footnote{
$f(t)=(1-r^t)-\frac{tr^t(1-r)}{r}-\frac{.5t(t-1)(1-r)^2r^{t-1}}{r}$ is a monotonically increasing function of time flattening out at a maximum value of 1, $g(t)=t^{2(L+1)}r^{t-1}(dr/ut)^\mc{D}$ is the time and distance dependent function which goes asymptotically to zero and $h_2=\frac{(d^2-dx^2)(d-x^2)}{(d^4-1)}\frac{(L-1)N_d^{2L+1}}{L^L(L-2)^{2L+1}}$ is an exponentially small constant w.r.t. time.}. The time-scale $T_2$ thus effectively sets an \emph{equilibration} time for the spread of correlations for the average stochastic unitary dynamics - reminiscent of the mixing time of quantum walks on the circle \cite{aharonov:2001}. Evaluating the R.H.S. of Ineq. (\ref{largetform1}) using the asymptotic values $r^{t\to\infty}=0,f(t\to\infty)=1,g(t\to\infty)=0$ one gets $\eta(\mc{D},t=\infty)/d^{L/2}<\sqrt{d}\mc{M}$ which is slightly larger, due to the lower estimates of $a(S,t)$, than the asymptotic value obtained by directly substituting the Haar average w.r.t. global unitaries ($U\in\mc{U}(\mc{H}_V)$) on the whole space giving
$\braket{\mc{R}_2^t(\hat{O}_p^{\otimes 2}),\hat{O}_q^{\otimes 2}\mc{T}_{V}}\to_{t\to\infty}\braket{\int dU U^{\otimes 2}(\hat{O}_p^{\otimes 2})U^{\dagger\otimes 2},\hat{O}_q^{\otimes 2}\mc{T}_{V}}$ leading to
$\eta(\mc{D},t=\infty)/d^{L/2}<(\{2(d-x^2)(d-y^2)\}^{1/2}/d^2)||\hat{O}_p||_2||\hat{O}_q||_2=:\mc{M}$ which holds for $\forall \{p,q\}\in V$.\\

\emph{Conclusions and Discussion}- We showed how average RMS correlations in our model of a LRQC depend on the defining probability distributions. Here, because the sequence of local regions picked and the unitaries were independently and identically chosen at each time step from the uniform measure, this model of a LRQC may be considered to be a particular example of a system undergoing general quantum Markovian dynamics \cite{poulin:2010} but one where the relaxation time scales with the system size resulting in no clustering of correlations. Correlated choices of local regions and/or local operations can infact model Non-Markovian random piece-wise quantum processes \cite{harrow:2009,zanardi:2013a}. The present results may be relevant to understand, for eg., the effect of local noise in an algorithm distributed over many qubits due to improper rotations and/or faulty spatial addressing of gates \cite{knill:1998} or quantum computation with bounded depth quantum circuits with intervening classical layers \cite{kalai:2014}. Generalizing these results to other random quantum circuit models to account for experimentally relevant situations would thus be important in the context of fault tolerant quantum information processing.

\emph{Acknowledgements.}- SS would like to thank Paolo Zanardi for his valuable comments and feedback. Thanks are also due to Lorenzo Campos Venuti and Iman Marvian for useful comments during a visit to USC and Joe Fitzsimons for pointing out reference \cite{kalai:2014}.

\bibliographystyle{apsrev4-1}
\bibliography{refs}
\textbf{~~~~~~~~~~~~~~~~SUPPLEMENTARY MATERIAL}\\
\section{Average over the circuits as ensemble maps}
Ineq.~(\ref{r1r2bound}) requires the average over the circuits of the square of the Frobenius norm of the commutator $[C^t(\hat{O}_p),\hat{O}_q]$ and may be expressed as,
\begin{align}
&\overline{\tr{\{[C^t(\hat{O}_p),\hat{O}_q]^\dagger [C^t(\hat{O}_p),\hat{O}_q]}\}}\nonumber\\
&=\overline{\tr\{(C^t(\hat{O}_p)\hat{O}_q-\hat{O}_qC^t(\hat{O}_p))^\dgr(C^t(\hat{O}_p)\hat{O}_q-\hat{O}_qC^t(\hat{O}_p))\}}\nonumber\\
&=\overline{\tr\{(\hat{O}_q^\dgr C^t(\hat{O}_p)^\dgr-C^t(\hat{O}_p)^\dgr\hat{O}_q^\dgr)(C^t(\hat{O}_p)\hat{O}_q-\hat{O}_qC^t(\hat{O}_p))\}}\nonumber\\
&=\overline{\tr\{(\hat{O}_qC^t(\hat{O}_p)-C^t(\hat{O}_p)\hat{O}_q)(C^t(\hat{O}_p)\hat{O}_q-\hat{O}_qC^t(\hat{O}_p))\}}\nonumber\\
&=\overline{2\tr\{\hat{O}_q^2C^t(\hat{O}_p)^2-C^t(\hat{O}_p)\hat{O}_qC^t(\hat{O}_p)\hat{O}_q\}}\nonumber\\
&=\overline{2\tr\{\hat{O}_q^2C^t(\hat{O}_p)^2-C^t(\hat{O}_p)\otimes C^t(\hat{O}_p)\hat{O}_q\otimes \hat{O}_q \hat{\mc{T}}_{V}\}}\nonumber\\
&=2\tr\{\hat{O}_q^2\overline{C^t(\hat{O}_p)^2}-\overline{C^t(\hat{O}_p)\otimes C^t(\hat{O}_p)}\hat{O}_q\otimes \hat{O}_q \hat{\mc{T}}_{V}\}\nonumber
\end{align}

where in the fifth line above we have used the cyclicity under trace and in the sixth line used the swap trick for trace of a product of operators $A,B\in\mc{B}(\mc{H}_V)$ as a trace of their tensor product times the swap operator on two copies of the same space i.e.: $\tr_{\mc{H}_V}(AB)=\tr_{\mc{H}_V,\mc{H}_{V'}}(A\otimes B~\hat{\mc{T}}_V)$. Also, $\overline{C^t(\hat{O}_p)^2}=\overline{U_{\textbf{S}}\hat{O}_pU_{\textbf{S}}^\dgr U_{\textbf{S}}\hat{O}_pU_{\textbf{S}}^\dgr}=\overline{U_{\textbf{S}}\hat{O}_p^2U_{\textbf{S}}^\dgr}=\mc{R}_1(\hat{O}_p^2)$
and $ \overline{C^t(\hat{O}_p)\otimes C^t(\hat{O}_p)}=\overline{U_{\textbf{S}}\hat{O}_pU_{\textbf{S}}^\dgr\otimes U_{\textbf{S}}\hat{O}_pU_{\textbf{S}}^\dgr}=\overline{U_{\textbf{S}}^{\otimes2}\hat{O}_p^{\otimes2}(U_{\textbf{S}}^\dgr)^{\otimes2}}=\mc{R}_2^t(\hat{O}_p^{\otimes2})$.

\section{The action of the maps $\mc{R}_1$ and $\mc{R}_2$}
 The superoperator $\mc{R}_1$ acting on an operator $\hat{O}_p^2$ acting at site $p$ is obtained by expanding Eq.~(\ref{ensemblemaps}):
\begin{align}
&\mc{R}_1(\hat{O}^2_p)=\frac{1}{L}\sum_{i=1}^{L}\int dU_{i,i+1} U_{i,i+1}~\hat{O}^2_p~U^\dgr_{i,i+1}\nonumber\\
&=r\hat{O}^2_p+\frac{1}{L}\int dU_{p-1,p} U_{p-1,p}\hat{O}^2_pU^\dgr_{p-1,p}+\frac{1}{L}\int dU_{p,p+1} U_{p,p+1}\hat{O}^2_pU^\dgr_{p,p+1}\nonumber\\
&=r\hat{O}^2_p\otimes\openone_{p\pm1}+\frac{2}{L}\frac{\tr(\hat{O}_p^2)}{d}\openone_p\otimes\openone_{p\pm1}\nonumber\\
&=(r\hat{O}^2_p+w\openone_p)\otimes\openone_{p\pm1},~\text{where}~r=\frac{L-2}{L},~w=\frac{2\tr(\hat{O}_p^2)}{Ld}
\label{r1uniform}
\end{align}
establishing that $\mc{R}_1(\hat{O}^2_p)\in\text{Span}\{\hat{O}^2_p\otimes\openone_{p\pm1},\openone_p\otimes\openone_{p\pm1}\}$. 
This implies that the non-trivial support of the operand does not change under the map $\mc{R}_1$ with the number of iterations. Because $\mc{R}_1$ is also trace-preserving,  $\tr{\{\mc{R}_1(\hat{O}^2_p)\}}=\tr{\{\hat{O}^2_p\}}=r\tr{\hat{O}^2_p}+wd$, yielding $ w=\frac{(1-r)\tr{\{\hat{O}^2_p\}}}{d}$. Iterations of $\mc{R}_1$ give,
\begin{align}
&\mc{R}_1^t(\hat{O}^2_p)=r^t\hat{O}^2_p\otimes \openone_{V\backslash p}+\frac{1-r^t}{d}\tr{\{\hat{O}^2_p\}}\openone_p\otimes \openone_{V\backslash p}\nonumber\\
\end{align}
resulting in $\braket{\mathcal{R}_1^t(\hat{O}^2_p),\hat{O}_q^2}=\tr\{\hat{O}^2_p\}\tr\{\hat{O}^2_q\}~d^{L-2}$ which is independent of time.
Next, expanding Eq.~(\ref{ensemblemaps}) for $\mc{R}_2$ we get,
\begin{align}
&\mc{R}_2(\hat{O}^{\otimes 2}_p)=\frac{1}{L}\sum_{i=1}^{L}\int dU_{i,i+1} U^{\otimes2}_{i,i+1}~\hat{O}^{\otimes2}_p~(U^\dgr)^{\otimes2}_{i,i+1}\nonumber\\
\end{align}
where using the result for integration over tensor product representations of unitary groups $\int dU_{i,i+1} U^{\otimes2}_{p,p\pm1}~\hat{O}^{\otimes2}_p~(U^\dgr)^{\otimes2}_{p,p\pm1}=\frac{2\tr(\hat{O}_p^{\otimes2}\Pi_{p,p\pm1})}{d^2(d^2+1)}\Pi^+_{p,p\pm1}+\frac{2\tr(\hat{O}_p^{\otimes2}\Pi_{p,p\pm1})}{d^2(d^2-1)}\Pi^-_{p,p\pm1}$ with $\Pi^{\pm}=(\openone\pm\hat{\mc{T}}_{p,p\pm1})/2$ the projectors onto the totally symmetric and antisymmetric spaces of $\mc{H}_d^{\otimes2}$ respectively and simplifying we get for the one-step action,
\begin{align}
\mc{R}_2(\hat{O}_p^{\otimes 2})&=r\hat{O}^{\otimes 2}_p+\frac{2A}{L}\openone_V+\frac{B}{L}(\hat{\mc{T}}_{p,p+1}+\hat{\mc{T}}_{p,p-1}),
\label{onestepr2}
\end{align}
which implies that $\mc{R}_2(\hat{O}_p^{\otimes 2})\in\text{Span}(\hat{O}_p^{\otimes 2},\openone,\hat{\mc{T}}_{p,p+1},\hat{\mc{T}}_{p,p-1})$.
In general for a fixed operator $\hat{O}_p^{\otimes2}$ the image $\mc{R}^t_2(\hat{O}_p^{\otimes 2})$ lies in a $(2^{|V|}+1)$-dimensional space of operators $\hat{O}_p^{\otimes2}\cup \hat{\mc{T}}_{S\subset V}$ where the $\hat{\mc{T}}_S$ are swap operators acting on subsets $S\subset V$ that swap the corresponding copies in the doubled Hilbert space $\mc{H}_V\otimes \mc{H}_{V'}$. For example, $\hat{\mc{T}}_{S=\phi}=\openone_{V}$, $\hat{\mc{T}}_{S=\{1\}}=\hat{\mc{T}}_{1}\otimes \openone_{\{V\backslash 1\}}$ etc. However for the specific model of LRQCs considered here there is an exponential reduction in the size of the invariant subspace to swap operators with support only on contiguous subsets $W\subset V$. We note that the set of operators $\hat{O}_p^{\otimes 2}, \hat{\mc{T}}_S,S\in W$ form an invariant subspace under the action of the map $\mc{R}_2$, i.e., $\mc{R}_2^t(\hat{O}_p^{\otimes 2})\in\text{Span}~(\hat{O}_p^{\otimes 2},\hat{\mc{T}}_{S\in W})$. One can then evaluate $\mathcal{R}_2^n(O_p^{\otimes 2})$ by writing down a difference equation,
\begin{align}
\mc{R}_2^{t+1}(\hat{O}_p^{\otimes 2})-r\mc{R}_2^t(\hat{O}_p^{\otimes 2})&=\frac{2A}{L}+\frac{B}{L}(\mc{R}_2^t(\hat{\mc{T}}_{p,p+1})+\mc{R}_2^t(\hat{\mc{T}}_{p,p-1})),
\label{Riter}
\end{align}
Iterating which for $n=t-1,t-2,t-3,...,0$ and adding those up we get,
\begin{widetext}
\begin{align}
\mc{R}_2^{t}(\hat{O}_p^{\otimes 2})&=r^{t}(\hat{O}_p^{\otimes 2})+\frac{2A}{L}\sum_{i=0}^{t-1}r^i+\frac{B}{L}(\mc{R}_2^{t-1}(\hat{\mc{T}}_{p,p+1})+r\mc{R}_2^{t-2}(\hat{\mc{T}}_{p,p+1})+...+r^{t-1}\hat{\mc{T}}_{p,p+1})+\frac{B}{L}(\mc{R}_2^{t-1}(\hat{\mc{T}}_{p,p-1})+...+r^{t-1}\hat{\mc{T}}_{p,p-1})
\label{r2long}
\end{align}
\end{widetext}

\section{Spatio-Temporal behavior of $a(S,t)$}
\begin{figure} 
\centering
\includegraphics[width=.9\columnwidth,height=5cm]{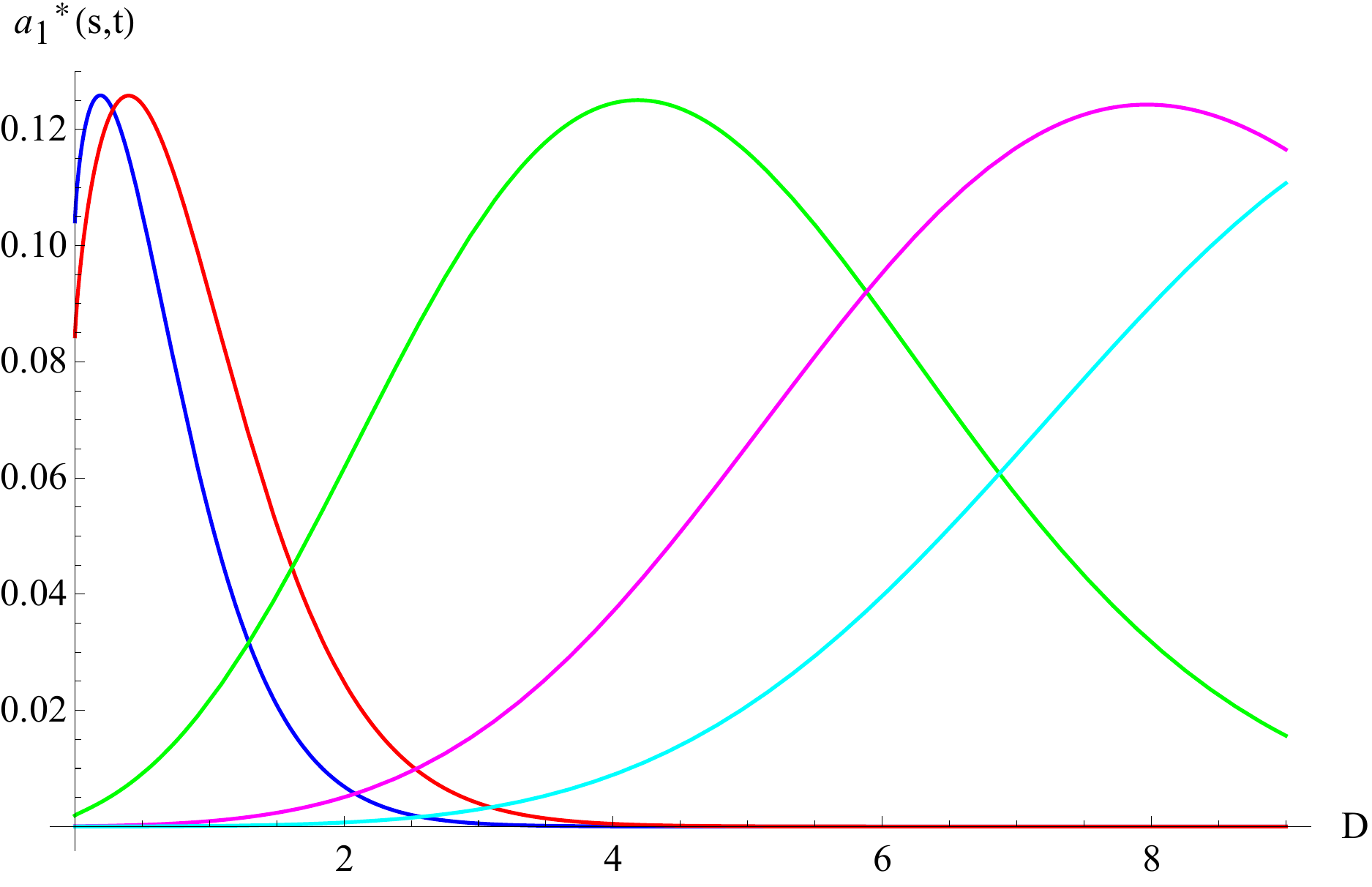}
\caption{(color online) Spatial dependence of $a^*(S,t)$ at various fixed instants of time for a circular chain with $L=10$. The blue ($t=10<T_1$) curve shows monotonic decrease of $a^*(S,t)$ for discrete various of $D=0,1,2,...,(L-1)$. The Red ($t=20$), Green ($t=200$) and Magenta ($t=380$) curves are for times $T_1<t<T_2$ which show the a maxima for intermediate values of $D$, $0<D<D_{\text{max}}=(L-1)$. The Cyan ($t=500>T_2$) curve shows monotonic increase with $D$. The actual values have been amplified by various factors (Blue:$\times 10^0$, Red:$\times 10^1$, Green:$\times 10^{19}$, Magenta:$\times 10^{37}$, Cyan:$\times 10^{49}$) to set comparable orders of amplitude in order for their features to be explained on the same plot.}
\label{ast}
\end{figure}

The behavior of $a_{1,2}^*(S,t)$ with increasing distance $D=\text{Dist}(S,S_{L,R})$ from the respective fiducial nodes show two distinct regimes. For $t<T_1=\frac{r}{du}=(1+1/d^2)(L-2)=O(L)$ there is monotonic decrease of $a_{1,2}^*(D,t)$ with distance while for $t>T_2=\frac{e(L+1)(L-2)(d^2+1)}{d}=O(L^2)$ there is monotonic increase of $a_{1,2}^*(D,t)$ with $D$, Fig.~(\ref{ast}). We use this fact in extracting the leading contributions to the dynamical bound. \\

\begin{figure} 
\centering
\includegraphics[width=\columnwidth]{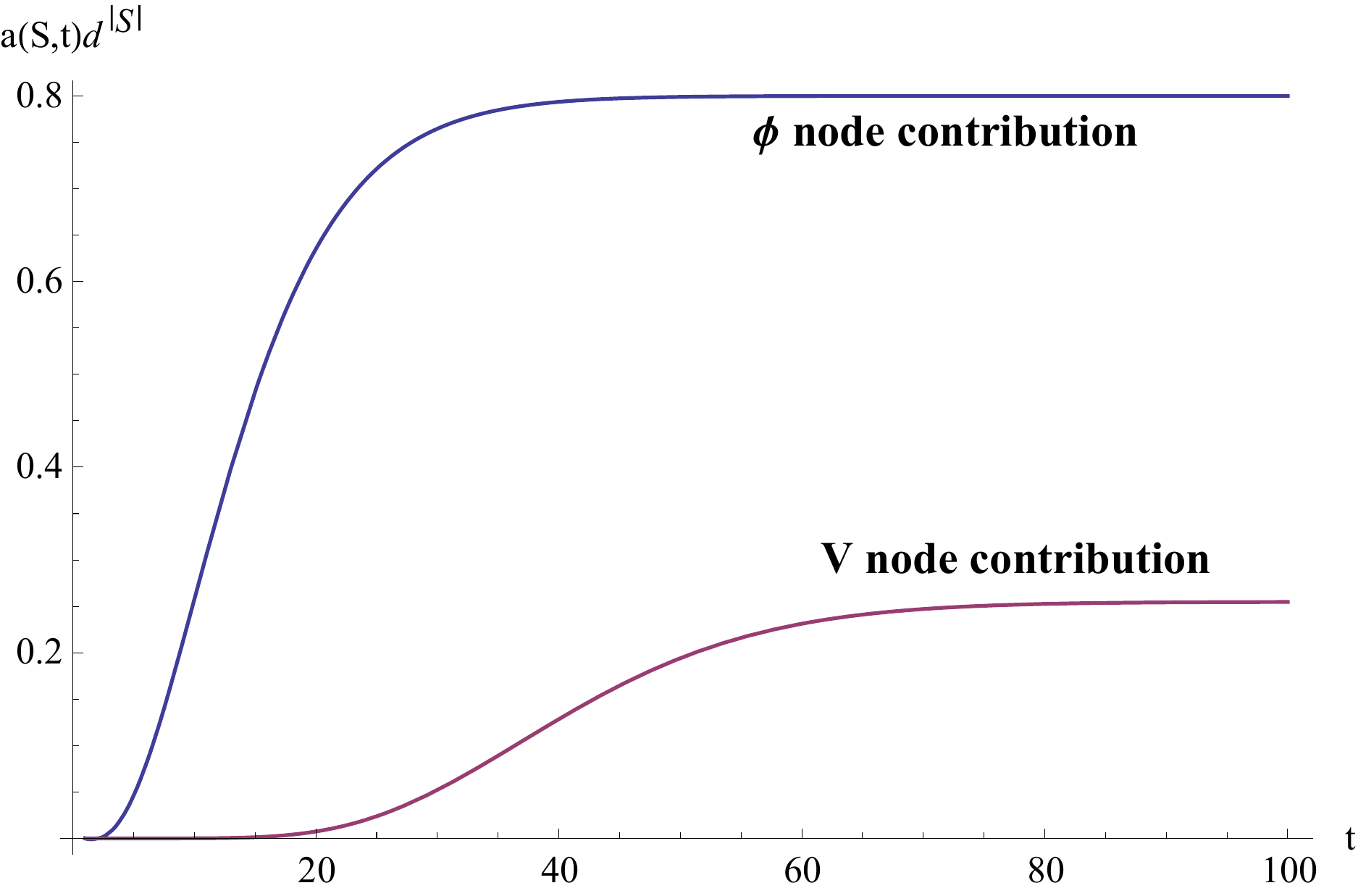}
\caption{(color online) Asymptotic contributions from the sink nodes for a chain with $L=10$. The contributions from the $S=\{V\}$ node is multiplied by a factor of $100$ to bring it to a comparable scale to be shown on the same graph.}
\label{asymptcontri}
\end{figure}

\begin{align}
&\sum_{S\in W}a(S,t)Q_S=\sum_{\overset{S\in W}{S\cap\{q\}=q}}a(S,t)Q_{1_S}+\sum_{\overset{S\in W}{S\cap\{q\}=\phi}}a(S,t)Q_{2_S}\nonumber\\
&=\sum_{\overset{S\in W}{S\cap\{q\}=q}}a(S,t)Q_{1_S}+(\sum_{S\in W}a(S,t)Q_{2_S}-\sum_{\overset{S\in W}{S\cap\{q\}=q}}a(S,t)Q_{2_S})\nonumber\\
&=\sum_{S\in W}a(S,t)Q_{2_S}-\sum_{\overset{S\in W}{S\cap\{q\}=q}}a(S,t)\Delta_S,~~\Delta_S=(Q_{1_S}-Q_{2_S})
\end{align}

 by keeping the most significant contributions for both. For the distance independent term $\sum_{S \in W}a(S,t)Q_{2_S}$ for short times $a(S,t)\simeq 1$ for the fiducial nodes whereas $a(S,t)=O(1/L)$ for all other ones whereas for long times $a(S,t)$ is exponentially small for all nodes except for the sink nodes $S=\{\phi\}$, Fig.~(\ref{asymptcontri}) . In both time-regimes we keep only the leading order contributions in time for the distance dependent term $\sum_{\overset{S\in W}{S\cap\{q\}=q}}a(S,t)\Delta_S$.

\end{document}